\documentstyle[preprint,revtex]{aps}
\begin{document}
\draft
\begin{title}
Boson-fermion model beyond mean-field approximation.
\end{title} 
\author{A.S. Alexandrov}
\begin{instit}

Department of Physics, Loughborough University of Technology, Loughborough LE11 3TU, U.K.

\end{instit}
\begin{abstract} 

\noindent A model of hybridized bosons and fermions is studied beyond the 
mean field approximation. The divergent  boson self-energy at zero temperature
makes the 
Cooper pairing of fermions impossible.
 The frequency and momentum dependence of the 
self- energy and the condensation  temperature $T_{c}$ 
 of initially localized
bosons  are  calculated analytically.  
  The value of the boson
condensation temperature $T_{c}$ is 
 below $1K$ which rules out the boson-fermion model with the 
initially localized bosons as a phenomenological explanation of 
high-temperature superconductivity. The intra-cell density-density fermion-boson interaction 
dominates in the fermion self-energy. The model represents a  normal 
metal  with strongly damped  bosonic excitations. The latter play
 the role of normal impurities.
 
\end{abstract} 


\narrowtext

{\bf 1. Introduction}

\noindent Many superconducting and normal state properties of perovskites 
 favor a charged $2e$ Bose liquid of small bipolarons
  as a plausible microscopic 
model of their ground state\cite{ale}. In particular, Bose-liquid features are clearly 
verified by the $\lambda$-like specific heat near the 
transition\cite{aleran}, the 
characteristic shape of the upper critical field \cite{ale2}, by the 
'boomerang" behavior of $T_{c}$ and the London penetration depth with 
doping \cite{uem}, explained recently \cite{alel}.

In a multi-band system a mixture of bipolarons and  electrons is 
feasible, with bipolarons formed in a narrow band
(the bandwidth $W<<E_{F}$) and almost free fermions, 
with a large Fermi energy $E_{F}$. If they interact with each 
other exclusively via the density-density interaction the effect of 
fermions is  that of the screening of the Coulomb boson-boson 
interaction. Hence, an acoustic gapless  plasmon  mode is expected in the 
entire temperature regime including the superfluid state while the 
fermionic component remains normal \cite{ale5}. The Bose-Einstein 
condensation temperature is expected to be about that of an 
ideal Bose-gas.

On the other hand Friedberg and Lee \cite{lee}, Ranninger and 
collaborators \cite{ran} and several other authors \cite{bar}  studied  
 bosons $hybridized$ with fermions, a so called $boson-fermion$ 
$model$ (BFM). The  BFM has been motivated by the difficulty to accommodate
a stable mobile
bosonic field  because of the allegedly strong Coulomb repulsion. Then 
the underlying mechanism for superconductivity has been assumed to be through 
the reaction $e+e \rightarrow \phi \rightarrow e+e$ involving $virtual$ 
$2e$
boson $\phi$. Because of this 
transition, it has been claimed that "the zero momentum virtual bosons force the two $e$'s to have 
equal and opposite momenta, forming a Cooper pair" \cite{lee}. The 
studies carried out on BFM with initially localized bosons  also 
showed a superconducting ground state, "controlled by the condensation of 
the bosons and a concomitantly driven $BSC$-like state of the fermionic 
subsystem" with the $BCS$-like gap in the electron spectrum \cite{ran}.
 It has been claimed \cite{lee,ran,bar} that BFM gives  the possibility of 
achieving large values of critical temperature.

In this paper we study the boson-fermion model  beyond the  mean field 
approach, 
applied in ref. \cite{lee,ran,bar}, by taking into account the boson 
self-energy and the intra-cell density-density repulsive interaction.
 We provide a  rigorous proof that the  Cooper pairing
of fermions is  impossible at any value of the repulsion. The ground state of BFM is 
essentially the same as that of the boson-fermion mixture with the normal 
fermionic component discussed earlier
 by 
us \cite{ale5}. The boson energy spectrum, damping and the density 
of states as well as the  Bose-Einstein condensation temperature $T_{c}$  
are calculated. The intra-cell correlations  dominate in the 
fermion self-energy. The role of the hybridization interaction is shown 
to be negligible both for $T_{c}$ and for the fermion damping.

\vspace{1cm}

{\bf 2. No Cooper pairing.}

\noindent BFM is defined by the following Hamiltonian \cite{lee,ran}
\begin{equation}
H=\sum_{{\bf k},s=\uparrow,\downarrow}\xi_{\bf k}c^{\dagger}_{{\bf 
k},s}c_{{\bf k},s}+\sum_{\bf q}\omega_{0}({\bf q})b^{\dagger}_{\bf q}b_{\bf q}+
{v\over{\sqrt{N}}}\sum_{\bf q,k}\left(b^{\dagger}_{\bf q}c_{{\bf 
k}+{\bf q},\uparrow}c_{-{\bf k},\downarrow}+H.c.\right),
\end{equation}
where $\xi_{\bf k}$ is the fermionic energy with respect to the chemical 
potential $E_{F}$, $\omega_{0}({\bf q})=E_{i}-2E_{F} +q^{2}/2M$ is the bare boson 
energy with $E_{i}$ the energy level of the doubly occupied $2e$ sites.
The bare boson mass $M$ can be infinite for  initially 
localized bosons \cite{ran}. The 
boson-fermion hybridization interaction $v\simeq \Gamma^{2}/|U|$ is
 of the second order with respect to the 
single-electron interband hybridization $\Gamma$. BFM is applied if the 
attractive on-site interaction $U<0$, responsible for the boson formation
 and the Fermi energy $E_{F}$ are 
large compared with $\Gamma$, so $v<<<|U|,E_{F}$. $N$ is the number of sites (cells) in the 
normalized volume and $\hbar=c=1$.

In all real-life systems the Coulomb repulsion 
exists, $V_{c}>0$, which is fairly represented by the Hamiltonian
\begin{equation}
H_{c}={V_{c}\over{N}}\sum_{{\bf k,k',q},s,s'}\left[c^{\dagger}_{{\bf 
k+q},s}c^{\dagger}_{{\bf k'-q},s'}c_{{\bf 
k'},s'}c_{{\bf k},s}+2c^{\dagger}_{{\bf 
k+q},s}c_{{\bf k},s}b^{\dagger}_{\bf k'-q}b_{\bf k'}\right],
\end{equation}
which describes the intra-cell correlations. 
If we want to keep within the limits of this 
particular BFM we shall take $V_{c}=0$. However, the strong 
inequality $V_{c}>>v$ is normally satisfied.

The criterion for Cooper-pair formation in a Fermi liquid lies in the 
existence of a nontrivial solution to the linearized BCS equation 
\cite{abr}
\begin{equation}
\Delta (p)=-\int dp' V(p,p') G(p')G(-p')\Delta (p').
\end{equation}
Under this condition the  two-particle vertex part, Fig.1a, has a pole 
in the Cooper channel. One can identify $\Delta$ with the superconducting 
order parameter and the temperature $T_{x}$ at which the nontrivial 
solution to Eq.(3) appears as the superconducting transition temperature for 
fermions. $V(p,p')$ is defined in the sense that it cannot be divided 
into two parts by cutting two parallel fermion propagators $G(p)$; 
$p\equiv ({\bf k}, i\omega_{n})$ is the momentum and the fermionic 
Matsubara frequency $\omega_{n}=\pi T (2n+1), n=0,\pm1,\pm2,...$, so that 
$\int dp'\equiv T\sum_{{\bf k},n}$. In the leading order in $v$
the irreducible interaction $V(p,p')$ is given by Fig.1b
\begin{equation}
V(p,p')={v^{2}\over{N}}D_{0}(0,0)+{V_{c}\over{N}},
\end{equation}
where 
\begin{equation}
D_{0}({\bf q}, \Omega_{n})={1\over{i\Omega_{n}-\omega_{0}({\bf q})}}
\end{equation}
is the $bare$ boson temperature Green's function with $\Omega_{n}=2\pi T 
n$.  
The physical (i.e. renormalized) fermion $G({\bf k},\omega_{n})$ 
and boson $D({\bf q},\Omega_{n})$ Green's functions satisfy the sum rule
\begin{equation}
{2T\over{N}}\sum_{{\bf q},n}e^{i\Omega_{n}\tau} D({\bf 
q},\Omega_{n})={T\over{N}}\sum_{{\bf k},n}
e^{i\omega_{n}\tau}G({\bf 
k},\omega_{n})-n_{e},
\end{equation}
which determines the chemical potential $E_{F}$ of the system. Here 
$n_{e}$ is the  carrier density per cell and $\tau=+0$.

By the use of Eq.(4) and Eq.(5) the BCS equation is reduced to a
simple form
\begin{equation}
1=\left({v^{2}\over{N\omega_{0}(0)}}-{V_{c}\over{N}}\right) \int dp'  G(p')G(-p'),
\end{equation}
which can be readily solved by replacing the exact fermion propagator  
for the bare one, $G(p)\simeq (i\omega_{n}-\xi_{\bf k})^{-1}$. Then the 
critical temperature is given by (for $V_{c}=0$)
\begin{equation}
T_{x}\simeq 1.14 E_{F}exp\left (-{\omega_{0}(0)\over{v^{2}N(0)}}\right),
\end{equation}
where $N(0)\simeq 1/E_{F}$ is the density of fermionic states. This 
is a 
mean-field result, which led several authors \cite{lee,ran,bar} to the
conclusion that BFM represents a high temperature superconductor if $\omega_{0}(0)$ is 
low enough. However, one has  to recognize that the bare boson 
energy $\omega_{0}(0)$ has no physical meaning and hence the expression 
Eq.(8) 
is meaningless. On the other hand the "physical" (i.e. renormalized) zero 
-momentum boson 
energy $\omega(0)$ is well defined. 
It should be positive 
or zero  according to  the sum rule, Eq.(6) in the renormalized model: 
\begin{equation}
\omega(0)=\omega_{0}(0)+\Sigma_{b}(0.0)\geq 0.
\end{equation}
Here $\Sigma_{b}({\bf q},\Omega_{n})$ is the boson self-energy given by 
Fig.2 \cite{ref} as
\begin{equation}
\Sigma_{b}({\bf q},\Omega_{n})= - {v^{2}\over{N}}T\sum_{{\bf k'},n'}
G({\bf k'}+{\bf q},\omega_{n'}+\Omega_{n})G(-{\bf k'},-\omega_{n'}).
\end{equation}
This equation is almost exact. The sixth and higher orders in $v$ 
"crossing diagrams", Fig.2b are as negligible as $v/E_{F}<<1$.
 It 
allows us to express the nonphysical bare energy $\omega_{0}(0)$ via the 
"physical" $\omega(0)$ as
\begin{equation}
\omega_{0}(0)=\omega(0)+{v^{2}\over{N}}\int dp'  G(p')G(-p').
\end{equation}
Then the BCS equation, Eq.(3) takes the following form
\begin{equation}
{\omega(0)\over{\omega(0)+({v^{2}/N})\int dp'  
G(p')G(-p')}}=-{V_{c}\over{N}} \int dp'  G(p')G(-p').
\end{equation}
It has no solution because $\omega(0)\geq 0$ and $V_{c}> 0$. 
Therefore we conclude that there is no pairing of fermions $(\Delta=0)$ 
in the boson-fermion model at any temperature in sharp contrast to the 
mean-field result of Ref.\cite{lee,ran,bar}. This conclusion is exact 
because no assumptions have been made as far as the fermion Green's function 
$G(p)$ is concerned. In particular, the taking into account of the fermion self- 
energy due to the hybridisation interaction $v$ or due to the fermion-fermion and 
fermion-boson repulsion (see below) does not affect our conclusion.

 One can 
erroneously believe that the renormalized boson propagator $D(0.0)$ rather than 
the bare one $ D_{0}(0.0)$ should be applied in the expression for
the irreducible vertex $V(p,p')$ (Fig.1b, Eq.(4)), in 
which case the physical zero-momentum energy $\omega(0)$ would appear in the 
expression for $T_{x}$, Eq.(8). This is incorrect because replacing 
$D_{0}$ for $D$ in the Cooper channel, Fig.1a, leads to a $double$ 
$counting$ of the hybridization interaction.  The same Cooperon 
diagram is responsible for the boson self-energy, Fig.2a. In other words 
the bare bosons only contribute to the Cooper channel. They are never 
condensed ($\omega_{0}(0) >0$ for any value of $v$) and, therefore cannot 
induce the superconducting state of the fermionic subsystem.  
From a pedagogical point of view it is 
interesting to note that a similar "double counting" problem appears in 
the calculation of the response function of condensed charged bosons. As 
has been 
discussed in Ref.\cite{alebeer} one should use the free particle 
propagator rather than the 
renormalized one to derive the textbook expression \cite{pin} for the boson 
dielectric response function. 

The authors of Ref.\cite{lee,ran,bar} 
applying the mean-field approach failed to recognize the 
divergence of the boson self-energy at zero temperature. It diverges 
logarithmically, so the bare boson energy is
 infinite at zero temperature and the pairing interaction ($\sim 
v^{2}N(0)/\omega_{0}(0)$) is zero. The divergent  boson self-energy 
fully compensates the divergent Cooperon diagram. Friedberg and Lee \cite{lee} discussed the 
 self-energy effect, missing, however,  the 
Fermi-distribution function in their out-of-place expression for the boson 
self-energy, which does not respect the Pauli 
principle (Eq.(1.15) of their paper).
As a result, they failed to 
notice  the  "infrared" collapse of their theory.

\vspace{1cm}

{\bf 3. Bose-Einstein condensation of strongly damped bosons}

While the fermionic subsystem remains normal at any temperature the 
bosons can be condensed at some finite temperature $T_{c}$. If their bare 
mass $M$ is sufficiently low, $T_{c}$ is given by the ideal Bose-gas 
formula \cite{ale5}, 
\begin{equation}
T_{c0}\simeq 3.3 {n_{B}^{2/3}\over{Ma^{2}}}, 
\end{equation}
where 
$n_{B}=(n_{e}-n_{F})/2$ is the boson  density per cell, $n_{F}$ is 
the fermion density, and $a$ is the lattice constant. 
 With their computer calculations of the boson 
damping Ranninger $et$ $al$\cite{ref} argue that the initially $localized$
 bosons with $M=\infty$
change over into free-particle-like propagating states as the temperature 
is lowered.
However, since the mean-field 
arguments are incorrect the
conclusion on the possibility of the Bose-Einstein condensation in BFM  
with $M=\infty$ is far from evident. 
 
 In this section we suggest an analytical calculation of
  the 
boson self-energy and show that despite  the $strong$ damping of the
long-wave bosonic excitations their condensation is possible. The 
critical temperature turns out to be rather low ($< 1K$).

The Bose-Einstein condensation temperature $T_{c}$ is given by 
the sum rule, Eq.(6) at $\omega(0)=0$
\begin{equation}
-{T\over{N}}\sum_{{\bf 
q},n}{e^{i\Omega_{n}\tau}\over{i\Omega_{n}+\Sigma_{b}(0,0)-\Sigma_{b}({\bf 
q}, \Omega_{n})}} =n_{B},
\end{equation}
By the use of the analytical properties of the boson self-energy the sum 
on the left-hand side is replaced by the integral as
\begin{equation}
\int_{0}^{\infty} {\rho (z)\over{exp(z/T_{c})-1}}dz=n_{B},
\end{equation}
where
\begin{equation}
\rho(z)= {1\over{\pi N}}\sum_{\bf q} {\gamma ({\bf 
q},z)\over{[z-\omega({\bf q},z)]^{2}+\gamma^{2} ({\bf 
q},z)}}
\end{equation}
is the boson  density of states. In the leading order in $v$ one can use 
the bare fermionic propagator in Eq.(10) to calculate 
 $\Sigma_{b}({\bf q},\Omega_{n}) $ as
\begin{equation}
\Sigma_{b}({\bf q},\Omega_{n})=-{v^{2}\over{N}}\sum_{\bf k}{tanh(\xi_{\bf 
k}/2T)+tanh(\xi_{\bf k+q}/2T)\over{\xi_{\bf k}+\xi_{\bf k+q}-i\Omega_{n}}}.
\end{equation}
 The analytical continuation to real frequencies is then
 \begin{equation}
\omega({\bf q},z)\equiv \Re \Sigma_{b}({\bf q},z)-\Sigma_{b}(0,0)
={z_{c}\over{4}}\int_{-\infty}^{\infty} dx tanh\left 
({qv_{F}x\over{4T_{c}}}\right) \left[ \ln 
{|x-1-z/qv_{F}|\over{|x+1-z/qv_{F}|}} 
+{2\over{x}}\right]
\end{equation}
for the real part, and
\begin{equation}
\gamma ({\bf q},z)\equiv \Im \Sigma_{b}({\bf q},z)=\pi z_{c} {T_{c}\over{qv_{F}}} \ln \left( 
{cosh{z+qv_{F}\over{4T_{c}}}\over {cosh{z-qv_{F}\over{4T_{c}}}}} \right)
\end{equation}
- for the damping. Here $z_{c}=v^{2}N(0)$ and $v_{F}$ is the Fermi velocity.
By the use of these equations we obtain  the following 
asymptotic behavior of the boson energy and of the damping in the long-wave  
$q<<q_{c}=4T_{c}/v_{F}$ and low energy $z<<qv_{F}$
 limit:
\begin{equation}
\omega(q)\simeq {q^{2}\over{2M^{*}}},
\end{equation}
\begin{equation}
\gamma \simeq z{\pi z_{c}\over{8T_{c}}},
\end{equation}
where the inverse  bosonic "mass" is determined by
\begin{equation}
{1\over{M^{*}}}={z_{c}v_{F}^{2}\over{6\pi^{2} 
T_{c}^{2}}}\sum_{n=1}^{\infty} {1\over{(2n-1)^{3}}}.
\end{equation}
Substitution of these expressions into Eq.(16) yields the square root 
asymptotic behavior of the boson density of states at low energies 
$z\rightarrow 0$ 
\begin{equation}
\rho (z) \sim \sqrt {z},
\end{equation}
which makes the integral in Eq.(15)  convergent and the Bose-Einstein 
condensation possible. However, the damping in this long-wave region is 
large. On the mass surface $z=\omega(q)$ we find
\begin{equation}
{\gamma\over {\omega(q)}}= {\pi z_{c}\over{8T_{c}}}\simeq 1
\end{equation}
because $T_{c} \leq z_{c}$ as we show below. Therefore the long-wave spectrum 
Eq.(20) 
as well as the bosonic mass $M^{*}$ due to hybridization have no physical meaning, and we 
reserve judgment whether or not a finite $T_{c}$ really signals the 
occurrence of superfluidity. 

The strongly damped part of the 
spectrum has a negligible weight in the total number of bosonic states 
due to a very small value of $q_{c}$ compared with the reciprocal 
lattice constant, $q_{c}<<1/a$. The inequality $q>>q_{c}$ is fulfilled
 practically in the whole Brillouin zone. In this case  by the use 
 of Eq.(18) and Eq.(19) we find
 \begin{equation}
 \omega(q)\simeq z_{c} \ln {q\over{q_{c}}},
 \end{equation}
 and
 \begin{equation}
 \gamma =z {\pi z_{c}\over{2qv_{F}}}.
 \end{equation}
 In this region the damping  is small, 
 $\gamma/\omega(q)<<1$, and the energy spectrum, Eq.(25) is well defined. 
 It is practically dispersionless, so the boson density of states is well 
represented by the $\delta$-function
 \begin{equation}
 \rho (z) \simeq \delta (z-z_{c}).
 \end{equation}
 Now the critical temperature is readily obtained from  Eq.(15) by the 
 use of Eq.(27) as 
 \begin{equation}
 T_{c}={v^{2}N(0)\over{\ln \left(1+1/n_{B}\right)}}.
 \end{equation} 
One can compare it with the condensation temperature $T_{c0}$, Eq.(13), 
determined with the 
finite bare boson mass $M\simeq |U|/W^{2}a^{2}$. Taking 
$n_{B}\simeq 1$ one obtains
\begin{equation}
{T_{c}\over{T_{c0}}}\simeq {\Gamma^{4}\over {E_{F}|U|W^{2}}}<<1,
\end{equation}
because in any realistic case $\Gamma \leq W$. As an example, if one 
believes \cite{ran,bar} that  localized bipolarons in $YBCO$ are associated with the 
$Cu-O$ chains and mobile single-particle states are associated with the
$Cu-O$ planes, the hydridization matrix element $\Gamma$ is 
proportional to  the 
chain-plane overlap integral and is clearly of the same order or even 
less than the intra-chain hopping integral ($\sim W$). As far as the polaronic 
reduction of the bandwidth is concerned,  by 
the use of the displacement canonical transformation and the Holstein 
model\cite{ale} one can readily show  that the reduction factor is 
precisely the same for both the bare bipolaronic band (the bandwidth 
$\sim1/Ma^{2}$)
and for the hybridized 
one (the bandwidth of the order of $v^{2}/E_{F}$). Moreover,
for intersite bipolarons and dispersive phonons the polaron orthogonality 
blocking of the (bi)polaron tunneling is less
significant than in the Holstein model 
as discussed recently by us \cite{alel}. At the same time the hybridization 
matrix element $\Gamma$  remains suppressed to the same extent  as 
in the dispersionless 
Holstein model\cite{tin}. Therefore, in general the  
orthogonality blocking (phonon overlap) 
reduces the ratio $T_{c}/T_{c0}$ even further. As a result, the relative 
 value of the 
critical temperature of the (localized) boson-fermion model is  small. 
The absolute value is very low as well. By taking 
$E_{F}\sim |U| \simeq 1eV$ and $\Gamma\leq 0.1 eV$ one estimates 
$T_{c}\simeq \Gamma (\Gamma^{3}/E_{F}|U|^{2}) \leq 1K$ which rules 
out BFM as an explanation of high temperature superconductivity.
Therefore the effect of  hybridization 
on the Bose-Einstein condensation  is  negligible.  As we show below the 
effect of  hybridization on the normal state fermion spectra is  
also negligible, compared with that of the   boson-fermion 
repulsion ($\sim V_{c}$).

\vspace{1cm}

{\bf 4. Fermion self-energy.}

The effect of hybridization on the normal state fermion spectrum in 
the BFM with initially localized bosons was 
discussed by several authors (see for example ref.\cite{ala}). It was 
shown  that the hybridization leads to a one-particle 
self-energy equivalent of the one of the marginal Fermi liquid. If the temperature is  above $T_{c}\sim 1K$ the 
fermion self-energy $\Sigma_{f}^{h}({\bf k},\omega_{n})$ due to  hybridization
  is well 
described by the following expression (see Fig.3)
\begin{equation}
\Sigma_{f}^{h}({\bf k},\omega_{n})=-{v^{2}\over{N}}T\sum_{{\bf 
k'},\omega_{n}'}{1\over{\left[i(\omega_{n'}+\omega_{n})-\omega(0)\right]\left[i
\omega_{n'}-\xi_{\bf k'}\right]}}.
\end{equation}
In this temperature range the finite boson bandwidth and the damping can 
be neglected while the use of the renormalized   
zero momentum boson energy  $\omega(0) > 0$ rather than $\omega_{0}(0)$ 
prohibits the violation of the sum rule, Eq.(6). The sum over 
frequencies is expressed as
\begin{equation}
\Sigma_{f}^{h}({\bf k},\omega_{n})={v^{2}\over{N}}\sum_{\bf 
k'}{coth{\omega(0)\over{2T}}-tanh{\xi_{\bf 
k'}\over{2T}}\over{i\omega_{n}-\omega(0)+\xi_{\bf k'}}}.
\end{equation}
Continued to the real frequencies this expression yields the 
following result for the imaginary part
\begin{equation}
\Im \Sigma_{f}^{h}(z)=sign(z)\pi z_{c}
\left[coth{\omega(0)\over{2T}}-tanh{\omega(0)-z\over{2T}}\right].
\end{equation}
Then the fermion lifetime due to hybridization is expressed using the sum 
rule as
\begin{equation}
 \Im \Sigma_{f}^{h}(z)=4\pi z_{c} 
{n_{B}(1+n_{B})\over{1+2n_{B}-tanh(z/2T)}}.
\end{equation}
If the temperature $T<<\omega(0)$ and the fermion  is far away from the 
Fermi surface 
($z>\omega(0)$) the lifetime is
\begin{equation}
\Im \Sigma_{f}^{h}=2\pi z_{c},
\end{equation}
which is the Fermi golden rule for spontaneous transitions to the empty local 
pair states. At the Fermi surface ($z=0$) the lifetime is proportional to 
the boson density
\begin{equation}
\Im \Sigma_{f}^{h}= 4\pi z_{c} {n_{B}(1+n_{B})\over{1+2n_{B}}},
\end{equation}
because the fermion needs some energy to annihilate.

We compare the hybridization lifetime, Eq.(35) with the damping due to  boson density 
fluctuations coupled with the fermion density, Eq.(2). As far as the 
direct repulsion between fermions is concerned (the first 
term in Eq.(2)), its contribution is negligible near the Fermi 
surface if $V_{c}<<E_{F}$, which is assumed here. The corresponding leading
contribution from $H_{c}$ to the fermion self-energy is  
then presented in Fig.4 and expressed as 
\begin{equation}
\Sigma_{f}^{c}({\bf k},\omega_{n})={4V_{c}^{2}\over{N}}T\sum_{{\bf 
q},n'}{\Pi_{b}({\bf 
q},\Omega_{n'})\over{i(\omega_{n}-\Omega_{n'})-\xi_{\bf k-q}}},
\end{equation}
where 
\begin{equation}
\Pi_{b}({\bf 
q},\Omega_{n})={1\over{N}}\sum_{\bf q'}{coth{\omega({\bf 
q'+q})\over{2T}}-coth{\omega({\bf q'})\over{2T}}\over{i\Omega_{n}-\omega({\bf 
q+q'})+\omega({\bf q})}}
\end{equation}
is the boson response function. There is only static response in  the limit $M\rightarrow \infty$ of 
initially localized bosons, so that
\begin{equation}
\Pi_{b}({\bf 
q},\Omega_{n})={n_{B}(1+n_{B})\over{T}} \delta_{\Omega_{n},0}.
\end{equation}
Substituting this expression into Eq.(36) we finally obtain
\begin{equation}
\Sigma_{f}^{c}(z)=i sign(z) 4\pi V_{c}^{2}N(0)n_{B}(1+n_{B}).
\end{equation}
The ratio of two lifetimes at the Fermi surface is given by
\begin{equation}
{\Im \Sigma_{f}^{h}\over{\Im 
\Sigma_{f}^{c}}}={v^{2}\over{V^{2}_{c}(1+2n_{B})}},
\end{equation}
which is about $10^{-4}$  for the appropriate values of $v$ and $V_{c}$.

It appears that the boson density fluctuations lead to an attractive 
interaction between two fermions, Fig.5, so that the total pairing 
potential is now
\begin{equation}
V(p,p')=-{v^{2}\over{N\omega_{0}(0)}}+{V_{c}\over{N}}-{4V_{c}^{2}\over{NT}}
n_{B}(1+n_{B})
 \delta_{\omega_{n},\omega_{n'}}.
 \end{equation}
 The BCS equation takes  the following form
 \begin{equation}
 \Delta(\omega_{n})=(\lambda -\mu)\pi 
 T_{c}\sum_{n'}{\Delta(\omega_{n'})\over{|\tilde {\omega}_{n'}|}}+
 4\pi V_{c}^{2}N(0)n_{B}(1+n_{B}){\Delta(\omega_{n})\over{|\tilde 
 {\omega}_{n}|}},
 \end{equation}
 where $\tilde {\omega}_{n}=\omega_{n}+sign(\omega_{n}) 4\pi 
 V_{c}^{2}N(0)n_{B}(1+n_{B})$ is the damped Matsubara frequency, 
 $\lambda=v^{2}N(0)/\omega_{0}(0)$, and $\mu=V_{c}N(0)$.
 Introducing a new order parameter $\tilde {\Delta}$ as
 \begin{equation}
 \tilde {\Delta}=\Delta(\omega_{n}) \left [1-{4\pi 
 V_{c}^{2}N(0)n_{B}(1+n_{B})\over{|\tilde {\omega}_{n}|}}\right],
 \end{equation}
 we obtain the same BCS equation, Eq.(7) as in the absence of any   
 density fluctuations
 \begin{equation}
 \tilde{\Delta}=(\lambda-\mu) \pi T_{c}\sum_{n'}
 {\tilde{\Delta}\over|\omega_{n'}|}.
 \end{equation}
 In this particular BFM it has only the trivial solution, 
 $\tilde{\Delta}=0$ because 
 $\lambda \rightarrow 0$ when $T\rightarrow 0$,  as explained above.
  As a result the density fluctuations of the 
 bosonic field have no effect on $T_{c}$, which is a textbook result 
 \cite{abr}. They play the same role in the BFM as 
 the normal impurities in  superconductors with no effect on the 
 critical temperature in accordance with the Anderson theorem.

{\bf Conclusion}

Our study of the boson-fermion model beyond the mean-field approximation 
has 
shown that this approximation, which predicts a $BCS$ -like 
fermionic superconductivity,
is qualitatively wrong. No Cooper pairing 
of fermions  due to their hybridization with the bosonic band 
is possible. The long-wave bosons are strongly damped and 
their condensation temperature is determined by the bare bosonic mass 
rather than by  hybridization. The fermion self-energy due to 
the  density-density coupling with bosons is larger by several orders of 
magnitude  than that due to hybridization.
BFM with initially localized bosons appears to be a  normal dirty metal  
where bosons play the 
role of  normal impurities.

Our results have a direct bearing upon the 
general problem of high-temperature superconductivity
via the exchange interaction. In a 
sharp contrast with the mean-field approach \cite{lee,ran,bar} the exact 
treatment of  the boson-fermion model leads to the 
conclusion that this model cannot provide a high value of $T_{c}$. 
Boson-fermion hybridization plays no role either in the $T_{c}$ value or 
in the fermion self-energy, which are determined by the bare effective 
mass of bosons and by the density-density 
fermion-boson coupling, respectively.

 Discussions with Sir Nevill Mott,
 Viktor Kabanov, with my colleges at the IRC and  Cavendish 
 Laboratory (Cambridge), and at  the Loughborough University Department of 
 Physics are highly appreciated.

\figure{ Two-particle vertex part for the Cooper channel ($a$) and 
irreducible  fermion-fermion interaction ($b$). }

\figure{Boson Green's function ($a$) and the lowest order "crossing" 
diagram ($b$).}

\figure{Hybridization contribution to the fermion self-energy.}

\figure{Fluctuation contribution to the fermion self-energy}

\figure {Effective attraction of fermions via boson density fluctuations.}

 \end{document}